\documentclass[a4paper,11pt]{article}
\usepackage{pos}
\usepackage{amsmath} 
\usepackage{graphicx} 
\usepackage{hyperref} 
\usepackage{url} 

\title{QCD in Language Models: What do they really know about QCD?}
\ShortTitle{QCD in LLMs}

\author*[a]{Antonin Sulc}
\author[b]{Patrick L.S. Connor}

\affiliation[a]{Lawrence Berkeley National Laboratory (LBNL),\\
  1 Cyclotron Road, Berkeley, CA 94720, USA}

\affiliation[b]{European Organization for Nuclear Research (CERN),\\
  Esplanade des Particules 1, 1211 Geneva 23, Switzerland}

\emailAdd{asulc@lbl.gov}
\emailAdd{patrick.connor@cern.ch}

\abstract{This study presents an analysis of modern open-source large language models (LLMs) -- including Llama, Qwen, and Gemma -- to evaluate their encoded knowledge of Quantum Chromodynamics (QCD). Through reverse engineering of these models' representations, we uncover the naturally idiosyncratic patterns in how foundational QCD concepts are embedded within their parameter spaces.
Our methodology combines targeted probing techniques and knowledge extraction protocols to assess the models' understanding of critical QCD principles like color confinement, asymptotic freedom, and the running coupling constant.
This work provides a tool for utilizing LLMs as an assistant in physics research, while also highlighting current limitations in their representation of advanced quantum field theory concepts that future model development should address.}

\FullConference{
2025 European Physical Society Conference on High Energy Physics\\
July 6-11, 2025\\
Marseille, France
}

\begin{document}
\maketitle

\section{Introduction}

The proliferation of large language models (LLMs) has marked a significant shift in numerous scientific domains. The availability of powerful open-weight models, trained on vast public data sets, presents a unique opportunity for the high-energy physics (HEP) community \cite{Barman2025, hallin2025foundation}. These models' ability to process and generate human-like language, which includes a substantial volume of scientific literature, raises crucial questions about their potential as research tools.

At their core, LLMs function as sophisticated next-token predictors. Given an input sequence, a model computes a probability distribution over its vocabulary for the subsequent token. This mechanism is the foundation of their ability to generate coherent text and to encode factual knowledge. This work, similar in spirit to other explorations like ChatQCD \cite{sulc2024chatqcd}, aims to address several key questions: How accurately do these models understand complex concepts in Quantum Chromodynamics (QCD)? How well can they relate different concepts? And ultimately, how can they be leveraged as reliable assistants in scientific research?

\section{Methodology: Measuring Knowledge with Perplexity}

To quantify an LLM's understanding, we employ a fundamental metric: \textbf{perplexity (PPL)}. Perplexity quantifies a model's surprise upon encountering a text sequence. A lower score indicates the model predicted the text accurately, suggesting that the information align with its training. Mathematically, perplexity is the exponential of the cross-entropy loss:
$$
\text{PPL}(W) = \exp\left( -\frac{1}{N} \sum_{i=1}^{N} \log P(w_i | w_{<i}) \right)
$$

We use this metric to design targeted tests, comparing the perplexity of different completions for a given prompt. For example, the statement \textit{"The strong force is mediated by \textbf{gluons}"} should yield a lower perplexity than \textit{"The strong force is mediated by \textbf{photons}"}. This method, known as probing~\cite{jelinek1998statistical}, is an established technique for querying the knowledge embedded within language models. 
\section{Results: Probing Foundational Concepts in QCD}

We applied our perplexity-based probing methodology to several open-weight LLMs, including models from the Llama 3 \cite{llama3modelcard}, Gemma \cite{team2025gemma}, and Qwen \cite{bai2023qwen} families, to evaluate their understanding of key numerical, categorical, and relational concepts in QCD and the Standard Model.

\subsection{Probing a Fundamental Constant: The Value of $\alpha_\mathrm{s}(M_\mathrm{Z})$}

\begin{figure}[h!]
    \centering
    \includegraphics[width=0.9\textwidth]{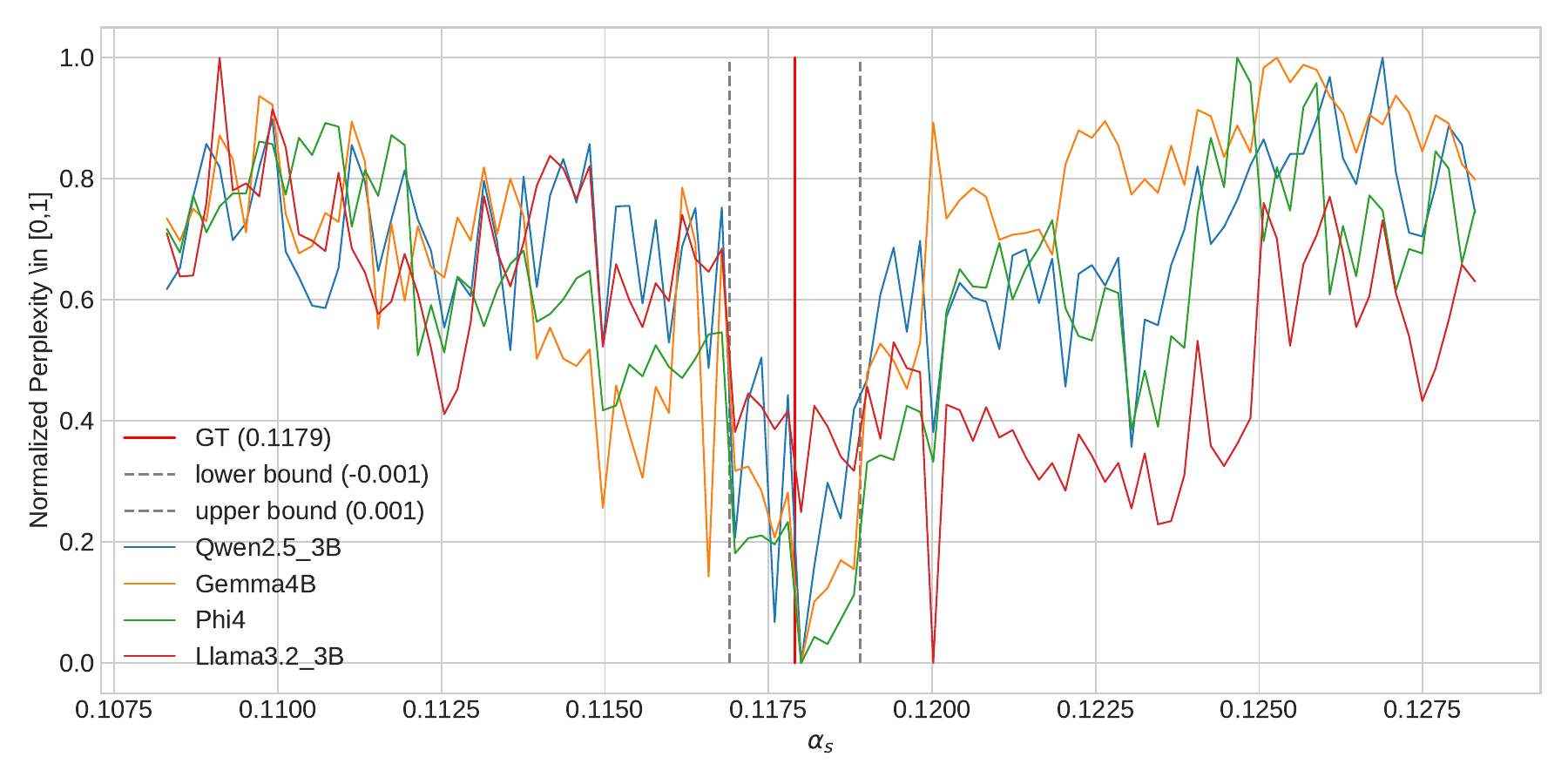}
    \caption{Perplexity as a function of the value of $\alpha_s$ in the prompt. The minimum for most models is observed near the accepted experimental value, indicating accurate numerical knowledge.}
    \label{fig:alphas}
\end{figure}

To test the models' grasp of numerical values, we probed their knowledge of the strong coupling $\alpha_s$. We provided prompts of the form \textit{"Strong coupling constant $\alpha_s$ is \{value\}"}, varying the value. The expectation is that perplexity will be minimized for the experimentally verified value of $\alpha_s(M_Z^2) \approx 0.118$ \cite{ParticleDataGroup:2024cfk}. As shown in Figure \ref{fig:alphas}, most models exhibit a distinct minimum in perplexity near the correct value, demonstrating they have encoded this constant with reasonable precision.

\subsection{Classifying Hadrons by Spin: Fermions vs. Bosons}

\begin{figure}[h!]
    \centering
    \includegraphics[width=0.9\textwidth]{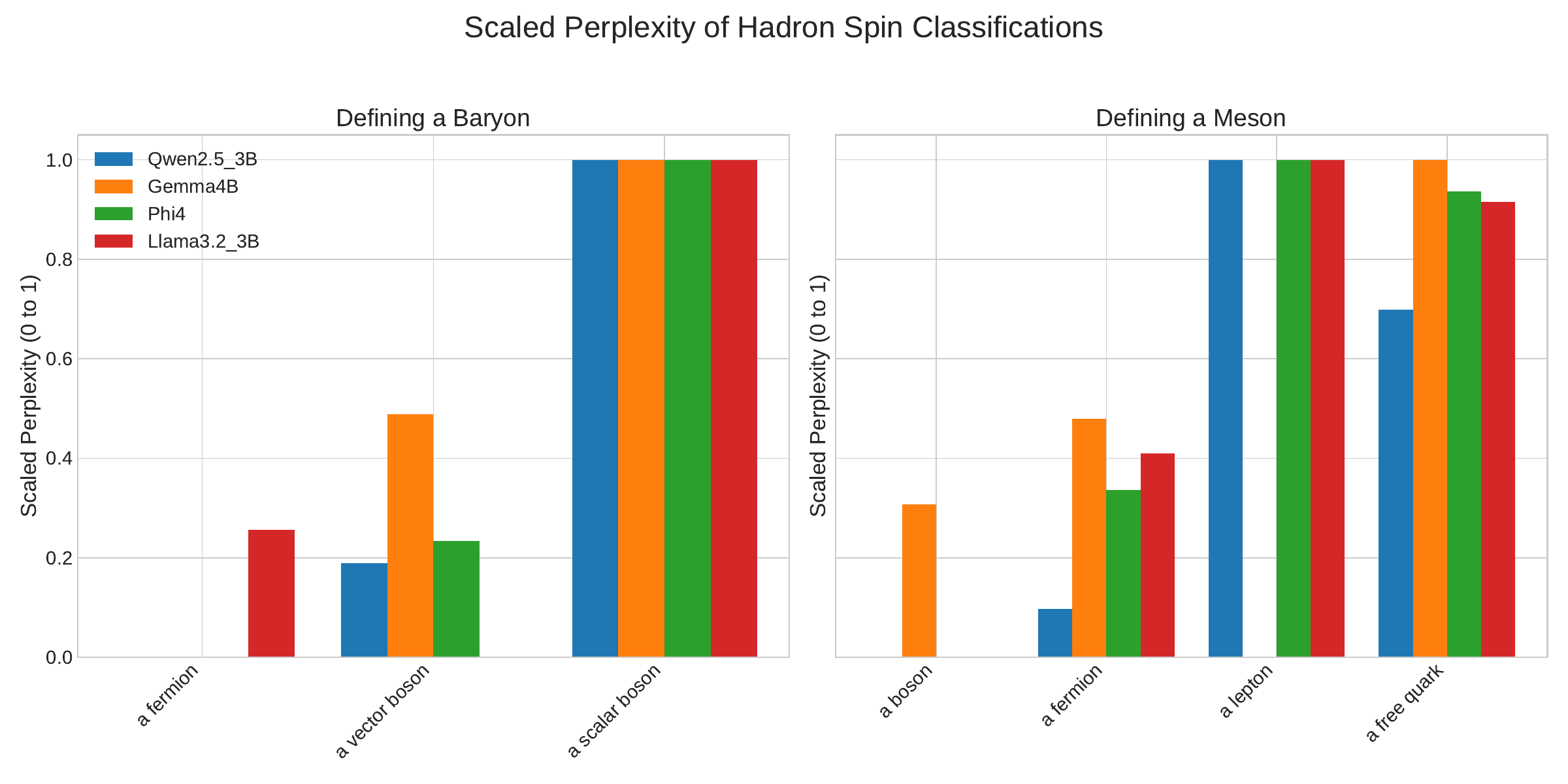}
    \caption{Scaled perplexity for hadron spin classification. Correct classifications (fermion for baryons, boson for mesons) yield the lowest perplexity.}
    \label{fig:spin}
\end{figure}

We investigated whether the models understand the spin-statistics theorem for composite particles using prompts like \textit{"Based on its total spin, a baryon is classified as \{classification\}"}. The results in Figure \ref{fig:spin} show that models correctly identify baryons as fermions and mesons as bosons, indicated by lower perplexity for correct classifications. However, some smaller models showed less certainty, highlighting a dependency on model size.

\subsection{Identifying Mediators of Fundamental Interactions}
\begin{figure}[h!]
    \centering
    \includegraphics[width=0.9\textwidth]{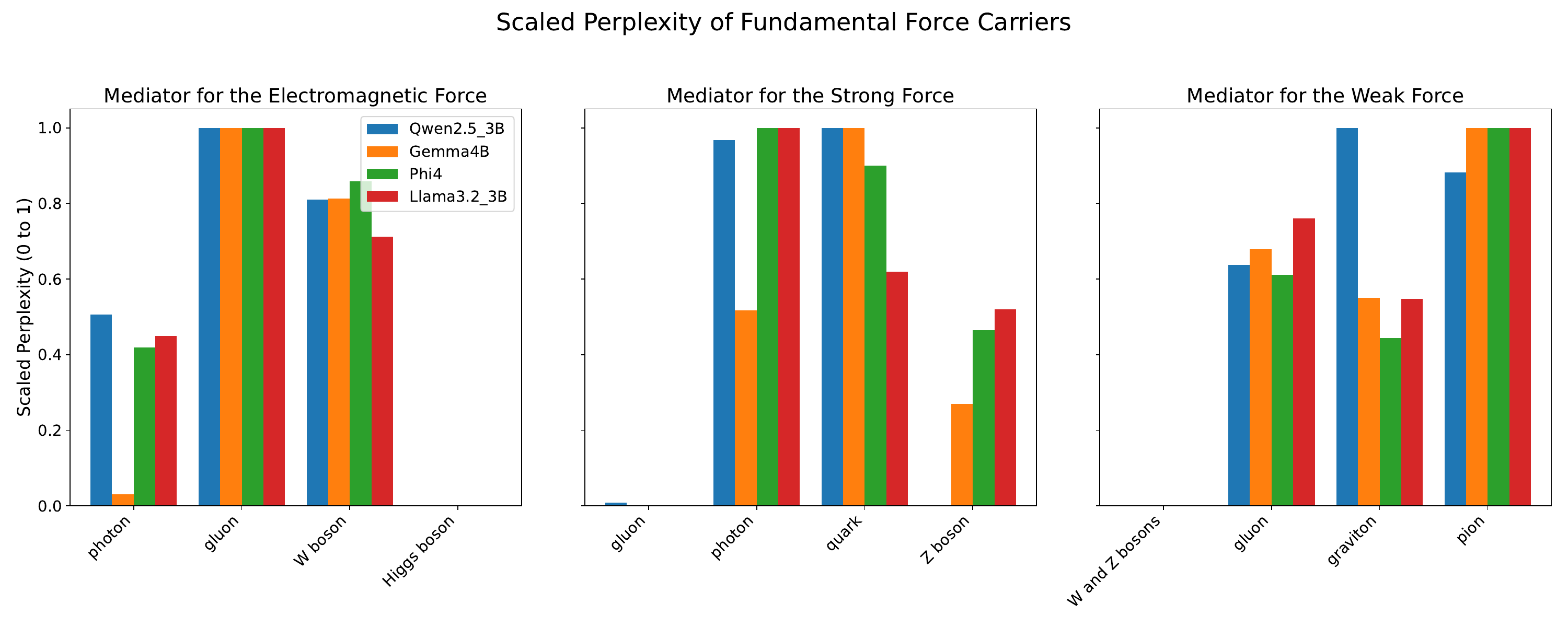}
    \caption{Scaled perplexity for prompts associating fundamental forces with their mediating bosons. Correct pairings consistently show the lowest perplexity.}
    \label{fig:carriers}
\end{figure}

We extended our analysis to the Standard Model's force carriers. As shown in Figure \ref{fig:carriers}, models correctly associated the strong force with gluons and the weak force with W/Z bosons. Surprisingly, results for the electromagnetic force were less clear-cut, suggesting that some associations be more weakly encoded than others.

\subsection{Context-Dependency of Quark Mass Knowledge}
\begin{figure}[h!]
    \centering
    \includegraphics[width=0.9\textwidth]{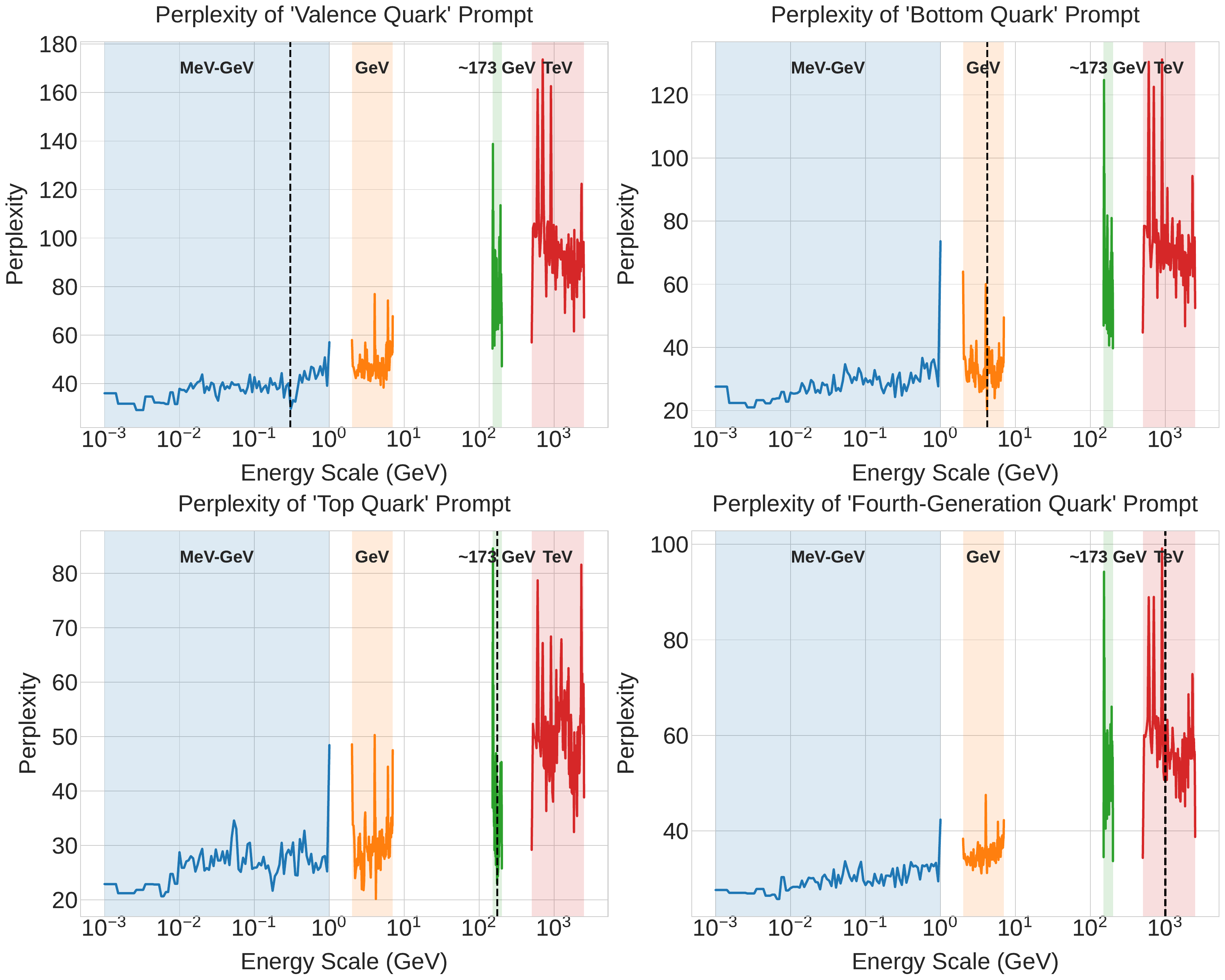}
    \caption{Perplexity scans for quark masses for the Llama3.2-3B model. The minimum perplexity is context-dependent, showing a correct peak for the top quark but less certainty for lighter quarks.}
    \label{fig:quarkmass}
\end{figure}

Lastly, we explored if a particle's mass is stored as a fixed value. We probed quark masses by scanning a wide range of energies. The results (Figure \ref{fig:quarkmass}) revealed a strong contextual bias. The perceived mass (point of lowest perplexity) shifted depending on the prompt's context, indicating that knowledge is stored as a dynamic representation rather than a static fact.

\section{Developing a Tool for Scientific Assistance}
While LLMs can make mistakes, their capabilities are sufficient to assist in scientific writing. Based on our findings, we are developing an AI companion to help researchers proof-check manuscripts. To demonstrate this capability, we have developed a public online validator tool, available at \href{https://huggingface.co/spaces/sulcan/EPS_HEP2025}{huggingface.co/spaces/sulcan/EPS\_HEP2025}.

To ensure the model evaluates statements from the correct scientific context and minimizes factual drift, the application employs a specific system prompt. This prompt frames the task as a validation against the Standard Model of particle physics, setting firm boundaries for what is considered established fact versus theoretical speculation. The system prompt is as follows:
\begin{quote}
\small
\texttt{System Framework: High Energy Physics Validation. This system evaluates all statements against the experimentally-verified Standard Model (SM) of particle physics. The baseline for physical reality is defined by Quantum Field Theory as applied through the SM's SU(3) x SU(2) x U(1) gauge structure.}\\
\texttt{The following principles are foundational and non-negotiable:}\\
\texttt{1. \textbf{Particle Content:} The fundamental constituents of matter and forces are exclusively the known quarks, leptons, gauge bosons, and the single Higgs boson of the Standard Model...}\\
\texttt{2. \textbf{Force and Mass Generation:} The electromagnetic and weak forces are unified within the electroweak theory. Fundamental particle masses arise solely through their coupling to the Higgs field...}\\
\texttt{3. \textbf{Conservation Laws:} All particle interactions must strictly conserve energy-momentum, charge, and lepton/baryon numbers...}\\
\texttt{4. \textbf{Beyond the Standard Model (BSM):} Concepts outside the SM (e.g., supersymmetry, string theory...) must be framed as theoretical, hypothetical, or speculative...}\\
\texttt{The system's function is to measure the perplexity of statements against this SM framework. Factual inaccuracies...will register as high-perplexity tokens.}
\end{quote}
This contextual steering is crucial for building a reliable scientific assistant, and our future work involves refining this approach for large-scale document analysis.

\section{Summary and Conclusion}
We have investigated the QCD knowledge encoded within publicly available LLMs. Our findings show that open-weight models perform reasonably well but still make mistakes, and no single model consistently outperforms others. Their knowledge is often context-dependent rather than static. Despite these limitations, LLMs are powerful enough to serve as assistants for scientific writing, especially when properly contextualized using system prompts. This work paves the way for their integration into the scientific workflow, while highlighting areas for future model improvement.

\end{document}